\documentclass[reprint,superscriptaddress,preprintnumbers,amsmath,amssymb,aps,prd,tightenlines,longbibliography,nofootinbib]{revtex4-2}

\usepackage{graphicx}
\usepackage{xcolor}
\usepackage[sort&compress]{natbib}
\usepackage{amsmath,amssymb,bm,bbm,slashed,subdepth,amsfonts}
\usepackage{xr-hyper}
\usepackage[colorlinks=true
,urlcolor=blue
,anchorcolor=blue
,citecolor=green
,filecolor=blue
,linkcolor=red
,menucolor=blue
,linktocpage=true
,pdfproducer=medialab
,pdfa=true
]{hyperref}
\usepackage{cleveref}
\usepackage{enumerate}
\usepackage{epsfig, subfigure}
\usepackage{setspace}
\usepackage{comment}
\usepackage{booktabs, tabularx}
\usepackage{units}
\usepackage{placeins}
\usepackage{multirow}
\usepackage{mathtools}
\usepackage[normalem]{ulem}
\graphicspath{{fig/}}

\newcommand{\be}{\begin{equation}}
\newcommand{\ee}{\end{equation}}
\newcommand{\bea}{\begin{equation}\begin{aligned}}
\newcommand{\eea}{\end{aligned}\end{equation}}

\definecolor{gesfblack}{rgb}{0,0,0}

\definecolor{gesfblue}{rgb}{0.08,0.42,0.76}

\definecolor{gesfgreen}{rgb}{0,1,0}

\definecolor{gesfgrey}{rgb}{0.5,0.5,0.5}

\definecolor{gesflanse}{rgb}{0.00,0.50,0.50}

\definecolor{gesfpurple}{rgb}{0.47,0.19,0.42}

\definecolor{gesfred}{rgb}{1,0,0}

\definecolor{gesfwhite}{rgb}{1,1,1}

\definecolor{gesfyellow}{rgb}{0.7,0.4,0.3}

\newcommand{\gsec}[1]{{\hypersetup{linkcolor=red}Sec.\,\ref{#1}\hypersetup{linkcolor=blue}}}
\newcommand{\gapp}[1]{{\hypersetup{linkcolor=red}App.\,\ref{#1}\hypersetup{linkcolor=blue}}}
\newcommand{\geqn}[1]{\hypersetup{linkcolor=blue}Eq.\,(\ref{#1})\hypersetup{linkcolor=blue}}
\newcommand{\gfig}[1]{{\hypersetup{linkcolor=violet}Fig.\,\ref{#1}\hypersetup{linkcolor=blue}}}

\definecolor{Orange}{cmyk}{0,0.61,0.87,0}
\definecolor{JungleGreen}{cmyk}{0.99,0,0.52,0}
\definecolor{OliveGreen}{cmyk}{0.64,0,0.95,0.40}
\definecolor{Brown}{cmyk}{0,0.81,1,0.60}
\definecolor{RoyalBlue}{cmyk}{0.71,0.53,0,0.12}
\definecolor{Gray}{cmyk}{0,0,0,0.40}
\definecolor{LightPink}{cmyk}{0.0,0.25,0,0}
\definecolor{LLightPink}{cmyk}{0.0,0.10,0,0}
\definecolor{LightBlue}{cmyk}{0.25,0,0,0}
\definecolor{LightGray}{cmyk}{0,0,0,0.2}

\allowdisplaybreaks[4]

\begin{document}

\title{Ultralight Dark Matter Spectrum with Gravitational Focusing and Doppler Motion}
 
\author{Shao-Feng Ge}
\email{gesf@stju.edu.cn}
\affiliation{State Key Laboratory of Dark Matter Physics, Tsung-Dao Lee Institute \& School of Physics and Astronomy, Shanghai Jiao Tong University, Shanghai 200240, China}
\affiliation{Key Laboratory for Particle Astrophysics and Cosmology (MOE) \& Shanghai Key Laboratory for Particle Physics and Cosmology, Shanghai Jiao Tong University, Shanghai 200240, China}

\author{Yong Tang}
\email{tangy@ucas.ac.cn}
\affiliation{University of Chinese Academy of Sciences (UCAS), Beijing 100049, China}

\author{Yue-Hui Yao}
\email{yaoyuehui21@mails.ucas.ac.cn}
\affiliation{University of Chinese Academy of Sciences (UCAS), Beijing 100049, China}
\affiliation{International Center for Theoretical Physics Asia-Pacific, Beijing, China}

\begin{abstract}
Ultralight dark matter can produce signals in experiments through its couplings with the Standard Model particles.
The resulting signal spectrum is determined by the phase-space distribution of dark matter, which encodes information about its production mechanism and the history of the local universe.
However, the local environmental effects can modify the observed spectrum and should be included in signal modeling to avoid biased parameter estimates.
In this work, we investigate how the gravitational focusing by Sun and the Doppler effect due to the detector’s motion modify the signal spectrum and affect inferring the underlying dark matter parameters.
We analytically estimate the signal significance and perform Bayesian analysis on simulated data to demonstrate their impact.
\end{abstract}

\maketitle

\section{Introduction}

Although the existence of dark matter (DM) receives
various strong evidences, its true nature is still unknown \cite{Bergstrom:2000pn, Bertone:2016nfn, Cirelli:2024ssz}. Among many possibilities, an interesting scenario is DM composed of ultralight bosons, including axion, dilaton, dark photon and other light fields motivated by physics beyond the Standard Model \cite{Peccei:1977hh, Wilczek:1977pj, Weinberg:1977ma, Kim:1986ax, Damour:1994zq, DiLuzio:2020wdo, Sikivie:2020zpn}.
These ultralight dark matter (ULDM) particles have
masses below a few electronvolts. At such low masses,
their de Broglie wavelengths can be astronomical with
wave effects to be significant on astrophysical scales
\cite{Marsh:2015xka, Hui:2016ltb, Hui:2021tkt, Ferreira:2020fam}.
In particular, the field spatial gradients give rise
to an effective pressure that can counteract gravity
and suppress structure formation on small scales,
potentially alleviating some of the small-scale
challenges to the conventional cold DM paradigm
\cite{Bullock:2017xww, Tulin:2017ara}.

Many theories predict that ULDM can couple to the
ordinary matter through interactions beyond the
Standard Model, which could produce observable
signatures in atomic clock
\cite{Arvanitaki:2014faa, Safronova:2017xyt},
pulsar timing array \cite{Liu:2019brz, Liu:2021zlt, Sun:2021yra, PPTA:2024mgh},
interferometer \cite{Pierce:2018xmy, Guo:2019ker, LIGOScientific:2021ffg, Heinze:2023nfb, Miller:2023kkd, Morisaki:2018htj, Grote:2019uvn, Yu:2023iog, Manita:2023mnc, Yu:2024enm, Liu:2025hwn, Yao:2025wfd, Hu:2026ato, Krauss:2026hrd},
and other experiments
\cite{Graham:2015ouw, Jiang:2021dby, Jiang:2023jhl, Xu:2023vfn, Li:2024zst, Fukuda:2025afi, Chigusa:2025rqs, Bao:2026gdw, Bogorad:2026hon},
as well as in astrophysical observations
\cite{Blas:2016ddr, Chen:2019fsq, Hardy:2022ufh, Arza:2023rcs, An:2023wij}.
In many of these experiments, the observable is a time series, which is then processed and analyzed in the frequency domain.
Since DM particles follow a velocity distribution, the ULDM field consists of modes with different velocities or different frequencies through the dispersion relation, with amplitude of each mode set by the fraction of DM particles in the corresponding phase-space volume \cite{Derevianko:2016vpm, Foster:2017hbq, Kim:2023pkx, Fukusumi:2023kqd}.
When the observation time is shorter than the field coherence time,
the signal spectrum cannot be resolved and appears as a monochromatic component, potentially accompanied by harmonics arising from the detector's periodic motion \cite{Amaral:2024tjg, Yao:2025vgy, Bouley:2026frx}.
However, when the observation time exceeds the coherence time, the spectral structure becomes resolvable and carries information about the phase-space distribution.
Analyzing this spectrum may help infer important properties such as the DM velocity dispersion and wind vector \cite{OHare:2017yze, Chang:2026zpo, Alcaraz:2026rdl}.

However, there are effects that can cause the observed
spectrum to deviate from the standard spectral model
predictions commonly adopted in the literature.
One such effect is gravitational focusing (GF) by the Sun, in which an incoming wave is scattered by the solar gravitational potential.
An earlier study \cite{Kim:2021yyo} shows that the
GF effect can modify the mean ULDM density along the Earth
orbit by a few percent.
The density, however, does not capture all the information.
For an observable that depends linearly on the ULDM field, the density is related to the total signal power, which is the integration of spectrum over frequency, and therefore does not retain the information encoded in the spectral shape.
 
In this work, we seek to extract the information encoded in the signal spectrum.
Specifically, we systematically investigate how
the GF and the Doppler effects affect the extraction
of the DM parameters from data. Our results demonstrate
that these effects are crucial for accurate signal
modeling and detection. 
 
The paper is organized as follows.
\gsec{sec:uldm and GF} introduces the field
description of ULDM and shows how the Sun's gravity
modifies the field to induce spatial variations
in its density.
In \gsec{sec:uldm spec}, we analyze the ULDM
spectrum in detail. Starting with the standard
spectrum where a flat spacetime and a static
detector are assumed, we then incorporate
the Doppler and GF effects in
\gsec{sec:Doppler effect} and \gsec{sec:GF},
respectively while the combined effect is presented
in \gsec{sec:total effect}. Guided by analytical
estimation of the signal significance in
\gsec{sec:snr criterion}, we test the impact
on parameter inference using simulated data
which is described in \gsec{sec:mock data}.
The results from Bayesian inference is summarized
and discussed in \gsec{sec:Bayesian infer}.
The final \gsec{sec:conclu_disc} gives our
conclusions and discussions. Through out this
paper, we use the natural units $c=\hbar=1$
but keep $G \neq 1$.

\section{Solar gravitational focusing for ultralight dark matter}
\label{sec:uldm and GF}

A scalar ULDM field in flat spacetime is described as a superposition of plane waves
\cite{Derevianko:2016vpm,Foster:2017hbq,Kim:2023pkx,Cheong:2024ose,Yao:2024fie},
\begin{equation}
  \phi(x)
\equiv
  \phi_0 \sum_{\mathbf{v}}
  \sqrt{\Delta_{\mathbf{v}}}
  \beta_{\mathbf{v}}
  e^{- i \left(\omega t - \mathbf{k}\cdot\mathbf{x} + \varphi_{\mathbf{v}}\right)} ,
\label{eq:phi plane waves}
\end{equation}
where $x \equiv (t,\mathbf{x})$ is a coordinate system
in the solar-system barycentric~(SSB) frame. 
While $m$ is the ULDM mass and $\mathbf{v}$ is the ULDM particle
velocity, $\omega \simeq m(1 + |\mathbf{v}|^2/2)$
is the energy
following the nonrelativistic dispersion relation.
Correspondingly, the ULDM momentum is
$\mathbf{k} \equiv m\mathbf{v}$. The prefactor
$\phi_0 = \sqrt{\bar{\rho}_0}/m$ is a constant
amplitude fixed by the DM energy density $\bar{\rho}_0$
and the phase space probability
$\Delta_{\mathbf{v}} \equiv F(\mathbf{v}) d \mathbf{v}$,
with $F(\mathbf{v})$ being the normalized DM velocity
distribution in the SSB frame, denotes the fraction of
particles within the velocity volume $d \mathbf{v}$.
Each plane wave has a random amplitude $\beta_{\mathbf{v}}$
from the Rayleigh distribution $\beta e^{- \beta^2 / 2}$
with unit scale parameter
and a phase $\varphi_{\mathbf{v}}$ with uniform distribution
over $[0, 2\pi)$.

It is convenient to separate the fast oscillation
$e^{- i m t}$
term from \geqn{eq:phi plane waves}
\begin{equation}
  \phi(x)
=
\sqrt{\frac 2 {m}}
\psi(t, {\bf x})   e^{-imt}
,
\label{def:psi}
\end{equation}
and define a complex field $\psi(t, {\bf x})$ to capture
the spatial dependence and the residual time dependence
besides the rest mass term $e^{\pm i m t}$ in \geqn{eq:phi plane waves}.
Since the Lorentz covariance is no longer
maintained in $\psi(t, {\bf x})$, we denote its 
time and spatial dependence explicitly rather
than using the combined form $x \equiv (t, {\bf x})$
that implies the Lorentz covariance.
With the definition in \geqn{def:psi},
the energy density of the field $\phi$ is given by
$\rho \simeq m^2 \phi^2 \simeq m|\psi|^2$, 
where the second ``$\simeq$" indicates averaging the
density over the rapid oscillations at Compton frequency
$f_C \equiv m/2\pi$, so that $|\psi|^2$ is interpreted
as the number density of ULDM particles \cite{Hui:2021tkt}.

With a non-relativistic velocity, $\psi$ varies
slowly over the coherence time
$\tau_c \equiv 2\pi/m\sigma^2$ and coherence length
$\lambda_c \equiv \sigma\tau_c$ where
$\sigma \sim 10^{-3}$ is the DM velocity dispersion around the Sun
\cite{Freese:2012xd}.
For light masses, these scales can be astronomical.
For example, the coherence length for
$m \sim 10^{-14}\,\text{eV}$
can be as large as the size of solar-earth system,
$\lambda_c \sim \mathcal{O}(\text{AU})$.

Under the influence of Sun's gravity, the evolution
of $\phi(x)$ follows the Klein–Gordon equation in curved space,
which reduces to the Schrödinger equation in the
non-relativistic limit \cite{Hui:2021tkt},
\begin{equation}
  i \frac{\partial}{\partial t} \psi(t, {\bf x})
=
\left(
- \frac{\nabla^2}{2m}
- \frac{GM_{\odot} m} r
\right) \psi(t, {\bf x}),
\label{eq:Schrödinger equation}
\end{equation}
where $M_{\odot}$ is the solar mass.
When passing by the solar system,
the plane wave solution for a wavy DM
would be focused by the solar gravitational
potential.
In the solar frame, an incoming wavy DM has positive
energy and hence would not form bound state in the
solar system. Instead, the incoming plane
wave would be deformed \cite{Kim:2021yyo} 
\begin{equation}
  \psi(t, {\bf x})
\propto
  e^{-i({\bf k}^2 t / 2 m  - \mathbf{k}\cdot\mathbf{x})}\chi_{\mathbf{k}}(\mathbf{x}).
\end{equation}
The exponential denotes the free plane-wave
solution and the scattering by the solar
gravitational potential is described by \cite{Landau:1991wop}
\begin{equation}
\label{def: chi}
  \chi_{\mathbf{k}}(\mathbf{x})
\equiv
  \Gamma(1 - i\eta) e^{\pi\eta/2}
  {}_1F_1(i\eta, 1, iz),
\end{equation}
where $z \equiv |{\bf k}| |{\bf x}| - {\bf k} \cdot {\bf x}
= 2 |{\bf k}| |{\bf x}| \sin^2\left(\theta/2\right)$
contains both the magnitudes of and opening angle $\theta$
between $\bf k$ and $\bf x$.
While $\Gamma(s)$ is the Gamma function, ${}_1F_1(a,b,c)$
is the confluent hypergeometric function. 
The parameter $\eta \equiv GM_{\odot}m/v$
characterizes the scattering strength.
With lower speed $v$, the DM wave would stay longer
around the Sun and hence the solar gravitational
focusing effect becomes more significant with
larger $\eta$.  

After gravitational focusing in the solar system,
the plane wave solution in \geqn{eq:phi plane waves}
would receive a correction term as $\chi_{\rm k}$.
Including the Sun's gravitational effect,
\geqn{eq:phi plane waves} becomes
\begin{equation}
  \phi(x)
=
  \phi_0 \sum_{\mathbf{v}}  \sqrt{\Delta_{\mathbf{v}}} \beta_{\mathbf{v}}
  e^{-i\left(\omega t - \mathbf{k}\cdot\mathbf{x} + \varphi_{\mathbf{v}}\right)}
  \chi_{\mathbf{k}}(\mathbf{x}).
\label{eq:phi hyp}
\end{equation}
The effect of GF is directly reflected in the field density contrast,
\begin{equation} \label{def:density contrast}
  \delta(x)
\equiv
  \frac{\rho(x) - \bar\rho_0}{\bar\rho_0}
=
  \frac{\phi^2(x)}{\phi^2_0} - 1 ,
\end{equation}
which also fluctuates over the coherence scales. Note that the density scales with the field squared ($\phi^2$) and includes a low-frequency stochastic part as well as a component oscillating at approximately twice of the Compton frequency $2 f_C$.
Using \geqn{eq:phi hyp}, the average density contrast is given by
\begin{equation} \label{def:avg density contrast}
  \bar\delta(\mathbf{x})
\equiv
  \langle\delta(x)\rangle
=
  \int {\rm d} \mathbf{v} F(\mathbf{v}) \left|\chi_{\mathbf{k}}(\mathbf{x})\right|^2 - 1,
\end{equation}
where $\langle\cdots\rangle$ denotes ensemble
average with
$\langle\beta_{\mathbf{v}}\beta_{\mathbf{v}'}\rangle=2\delta_{{\mathbf{v}}{\mathbf{v}}'}$.
Since the field is stationary, the average density
contrast has no explicit time dependence.

\begin{figure*}
\centering
\includegraphics[width=0.9\linewidth]{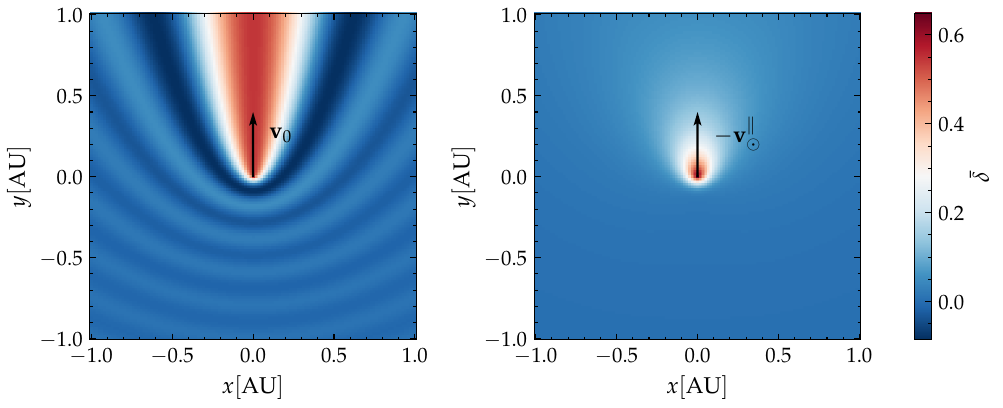}
\caption{Average density contrast for a
monochromatic plane wave along the indicated
direction ${\bf v}_0$ on the ecliptic plane
(left) and for the Maxwellian distribution
given in Eq.\,(\ref{eq:Maxwell dist}) (right).
The black arrows indicate the projected direction of the DM
wind.
}
\label{fig:avg density contrast}
\end{figure*}

It is instructive to start with the effect of GF on a
monochromatic plane wave with fixed velocity ${\bf v}_0$, i.e.
$F(\mathbf{v}) = \delta(\mathbf{v}-\mathbf{v}_{0})$.
In this case, \geqn{def:avg density contrast} reduces to
\begin{equation}
  \bar\delta(\mathbf{x}) 
=
  \left|\chi_{\mathbf{k}_{0}}(\mathbf{x})\right|^2 - 1 
\overset{\text{ds}}{=}
  \frac{2\pi\eta}{1 - e^{-2\pi\eta}}-1,
\end{equation}
where ``$\overset{\text{ds}}{=}$" denotes the value
along the downstream direction $z=0$ with $\theta = 0$,
which is independent of the distance $|\bf x|$
from the central object. Note that such monochromatic
plane wave can lead to an enhancement along the
downstream direction as a prominent feature.
The left panel of \gfig{fig:avg density contrast}
shows the density contrast in the $x$-$y$ plane for $m=2\times10^{-14}\,\text{eV}$
and $\mathbf{v}_{0} = [0, 10^{-3}, 0]$ under the
influence of solar GF. 
The wavelength is $0.4\,\textrm{AU}$ and comparable with the distance between Sun and Earth.
As illustrated in the figure,
the density contrast oscillates and becomes
increasingly distorted as it approaches the Sun.
In the downwind, a density wake eventually forms
with significantly enhanced magnitude. However,
the density in the upstream and in the perpendicular
directions are not that much affected.

However, DM in our galaxy is composed of
not just a single plane wave. After long evolution,
the DM particles should have reached thermal equilibrium
with Maxwell distribution \cite{Veltmaat:2018dfz,Schwabe:2021jne} in the galaxy frame. Since the solar system moves in the Milky Way galaxy, the
DM velocity distribution follows the boosted Maxwellian
distribution
\begin{equation}
  F(\mathbf{v})
= 
  \frac 1 {(2\pi \sigma^2)^{3/2}}
  e^{- \frac {(\mathbf{v} + \mathbf{v}_{\odot})^2} {2 \sigma^2}} ,
\label{eq:Maxwell dist}
\end{equation}
with $\mathbf{v}_{\odot}=[- 104.5, 34.9, 196.8]\,\text{km}\cdot\text{s}^{-1}$ being the solar velocity in our Milky Way galaxy\footnote{We define the SSB coordinate system such that the Sun is at the origin, the positive $x$-axis points toward the vernal equinox, the positive $y$-axis points toward the summer solstice, and the $z$-axis is normal to the Earth's orbital plane, forming a right-handed coordinate system.} and $\sigma=155.6\,\text{km}\cdot\text{s}^{-1}$ the velocity dispersion. 
In natural units, their magnitudes are $v_{\odot}=7.5\times10^{-4}$ and $\sigma=5.2\times10^{-4}$.
Here, the velocity values are taken from the Standard Halo Model (SHM) \cite{Drukier:1986tm,Freese:1987wu,Bozorgnia:2016ogo}.
The right panel of \gfig{fig:avg density contrast} shows $\bar\delta(\mathbf{x})$ in the ecliptic plane for $m=2\times10^{-14}\,\text{eV}$ after
doing the phase space integration with a Maxwellian
distribution.
Compared with the monochromatic case, a density
wake still forms behind the central object. However,
the originally dramatic enhancements in the downstream
with a single plane wave as shown in the left panel
now average out to a much more smooth profile. 
Especially, there is no oscillation any more over
the coherence length
$\lambda_c \approx0.8\,\textrm{AU}$ due to the
superposition of waves.
The resulting density pattern remains axisymmetric,
owing to the intrinsically preferred direction
${\bf v}_\odot$ of the velocity distribution.

\begin{figure}
\centering
\includegraphics[width=0.9\linewidth]{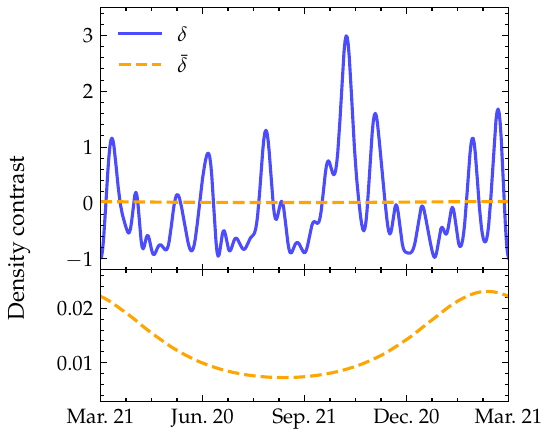}
\caption{Density contrast along the Earth orbit for $m=2\times10^{-14}\,\text{eV}$. The blue curve
in the upper panel shows the density contrast for a
specific field realization
while the orange one denotes the average density contrast.
}
\label{fig:density contrast}
\end{figure}

For ground-based detectors, their velocities relative to the DM background change annually due to the Earth's motion around the Sun.
When the Earth is moving towards the DM wind,
which happens around June 1st,
larger relative velocity would lead to larger event rate.
This is the annual modulation of DM direct
detection in the classical sense \cite{Freese:2012xd}.
With GF included, the average contrast shifts
with an amplitude of $\mathcal{O}(1\%)$ to peak
around March when the Earth is downstream of
the DM wind and reaching minimum around December
when it is upstream \cite{Lee:2013wza,Bozorgnia:2014dqa}.
\gfig{fig:density contrast} shows the annual
evolution of $\bar\delta(\mathbf{x})$ along the
Earth's orbit for $m=2\times10^{-14}\,\text{eV}$
and the velocity distribution in \geqn{eq:Maxwell dist},
alongside the density contrast for a specific realization.
The density exhibits large fluctuations that follow an exponential distribution,
while GF manifests as a small enhancement in the mean density.

Since the typical speed of galactic DM particles
exceeds the escape velocity of the solar system,
we only include the unbound states in current work.
However, bound states or substructures may still exist in
the solar neighborhood \cite{Banerjee:2019epw, Banerjee:2019xuy}.
While their relevance to observations depends strongly on their energy density, which is highly uncertain,
they possess properties distinct from the galactic DM halo considered here and can leave characteristic imprints in the signal spectrum.
For example, the gravitational atom may also form
around the Sun \cite{Paranjape:2026hkr}, in addition
to the background galactic halo. Since the ground
state of the atom is perfectly coherent, it will appear in the
Fourier domain as a spike and its amplitude grows
linearly with time.

\section{Spectrum with gravitational focusing and Doppler effect}
\label{sec:uldm spec}

\begin{figure*}[t]
\centering
\includegraphics[width=0.9\linewidth]{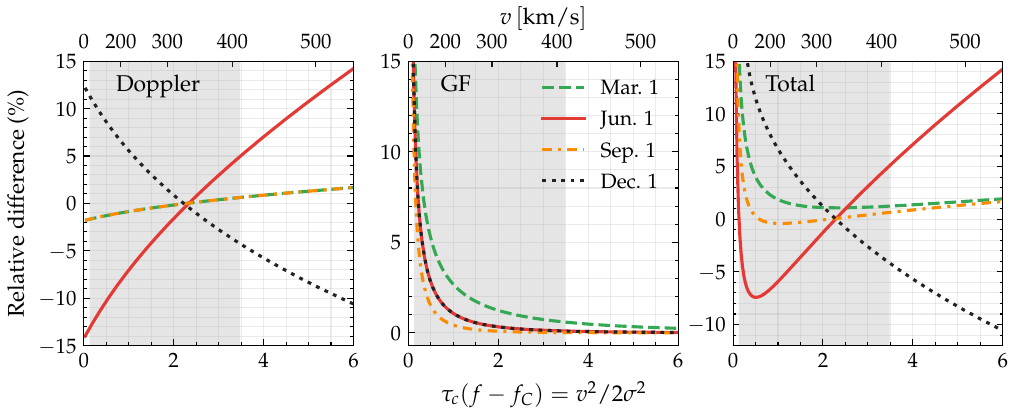}
\caption{Relative differences between the corrected spectra
and the standard spectrum on the first day of March (green dashed), 
June (red solid), September (orange dash-dotted), and December (black dotted)
with the Doppler effect only (Left), gravitational
focusing only (Middle), and total results with both effects combined (Right).
The gray shaded region indicates the frequency range where the spectral amplitudes exceed $\max[S_{\phi}]/2$.
While the lower horizontal axis for frequency
is independence of the velocity scale, we also put
the upper axis for velocity with the SHM scale
$\sigma = 155.6$\,km/s.}
\label{fig:relative diff}
\end{figure*}

We consider the case that the sensors inside the detector are linearly coupled to the field \cite{Morisaki:2018htj,Pierce:2018xmy,Gue:2024txz,Yao:2024hap}.
For example, a scalar field $\phi$ can couple
to the quark field $\psi$ through
$\propto \phi \overline{\psi}\psi$,
which would eventually lead to coupling to
the test masses (such as mirrors) in a laser
interferometer through the trace of their
energy-momentum tensor
\begin{equation}
    S = -\int m(\phi) \sqrt{-\eta_{\mu \nu} dx^{\mu} dx^{\nu}}.
\end{equation}
Consequently, the test masses tend to move
toward regions with lower effective mass,
leading to an acceleration along the sensitive
axis $\hat{\mathbf{n}}$ \cite{Morisaki:2018htj}
\begin{equation} \label{eq:gradient coupling}
\ddot x(t)\equiv \frac{d^2x}{dt^2} = -\kappa\alpha \hat{\mathbf{n}}\cdot\nabla\phi,
\end{equation}
where $\alpha$ is the coupling strength and $\kappa \equiv \sqrt{4\pi G}$ is the inverse of the Planck mass.

For a vector DM $V^{\mu} \equiv (V^0, \mathbf{V})$,
such as dark photon, which couples to test
masses via baryon number (B) or baryon-minus-lepton
number (B–L) \cite{Fayet:1990wx,Fayet:2016nyc}, the interaction is described by
\begin{equation}
  \mathcal L_{\text{int}}
=
- \epsilon e J^\mu V_\mu,
\end{equation}
where $e$ is the electric charge and $\epsilon$
is the coupling strength.
The concrete material property can be summarized into
the total dark charge $q$ of the mirror material.
Analogously to the electrodynamics, such interaction
mediated by the dark photon leads to an acceleration
along the sensitive axis $\hat{\bf n}$ \cite{Pierce:2018xmy}
\begin{equation}
\label{eq:dp_acc}
  \ddot x(t)
=
  \epsilon e \frac q M
  \hat{\mathbf{n}}\cdot\partial_t \mathbf{V}.
\end{equation}
where $M$ is the mirror mass. With both $q$ and $M$
being macroscopic properties, the charge-mass ratio
$q / M$ is an intrinsic property of the material.

For quadratic couplings, the matter effects
may also need to be taken into account with
ground-based experiments
\cite{Hook:2017psm, Banerjee:2025dlo, Gan:2025nlu, Bouley:2026frx}.
The presence of a large SM energy density can
alter the amplitude of ULDM field, leading to
either amplification or screening which depends
on the sign of the couplings and ULDM mass.
In particular, when the wavelengths of certain
field modes are comparable to the size of our
Earth, resonances may amplify these velocity
modes and thereby modify the spectrum.
For illustration and simplicity, we focus on
linear couplings in this work.

The signal or time series $s(t)$ recorded by the
detector is related to $\ddot x(t)$. In signal
processing, one usually analyze $s(t)$ in the
frequency domain. According to Percival's
theorem \cite{oppenheim2013discrete},
the integrated effect is
\begin{equation}
  \langle \ddot x(t)^2\rangle
\propto
  \langle \phi(t)^2\rangle
=
  \int^{\infty}_0 {\rm d} f S_{\phi}(f) ,
\end{equation}
where $S_{\phi}(f)$ is the ULDM signal power spectrum.\footnote{Strictly speaking, $S_{\phi}(f)$ should be replaced by $S_{\partial_t A_i}(f)$ and $S_{\partial_i\phi}(f)$, which are the spectra of the field derivatives and are not identical to the field spectrum $S_{\phi}(f)$. However, they do not differ essentially from $S_{\phi}(f)$ which we use instead for convenience.}
Therefore, variations in the density correspond to variations in the total signal power, which varies at the $\mathcal{O}(1\%)$ level as the detector orbits around the Sun.
As shown below, the GF and Doppler effects associated with the detector’s motion can modify the shape of $S_{\phi}(f)$ and thus affect the extraction of physical information.

While the exact signal spectrum $s(t)$ depends
on the detailed experimental setup and
coupling mechanism, it shares a common
structure, as demonstrated above. 
Without loss of generality, we use the scalar
case $s(t) = \phi(t)$ as the primary
example in what follows.

The signal power spectral density (PSD)
is defined as
\begin{equation}
\label{def:psd}
  S_{\phi}(f)
\equiv
  2 \frac{\langle|\tilde{\phi}(f)|^2\rangle}{T},
\end{equation}
where $\tilde{\phi}(f) \equiv \int {\rm d} t e^{i 2 \pi f t} \phi(t)$
is the Fourier transform
of $\phi(t)$ and $T$ is the signal duration.
Using \geqn{eq:phi plane waves} and \geqn{def:psd}, we can show that the ULDM spectrum for a flat space and a static detector is given by
\begin{equation}
\label{eq:standard spec}
  S_{\phi}(f)
=
  \frac{v}{f_C} \int {\rm d}\hat{\mathbf{v}}  F(\mathbf{v}),
\end{equation}
where $v=\sqrt{2(f/f_C - 1)}$  is the non-relativistic
velocity with $f / f_C \equiv \omega / m$ and
$\mathbf{v}=v\hat{\mathbf{v}}$ with $\hat{\mathbf{v}}$
being a unit vector to denote its direction.
We have set $\phi_0=1$ for convenience and
implemented $\langle\beta_{\mathbf{v}}\beta_{\mathbf{v}'}\rangle=2\delta_{{\mathbf{v}}{\mathbf{v}}'}$.
Note that the original phase space integration
element $d {\bf v} = v^2 d v d \hat{\bf v}$
has been converted to the frequency integration
with $v d v = d f / f_C$ since
$f = f_C (1 + v^2/2)$.

Thus the power contained in a given frequency interval is directly proportional to the number density of DM particles within the corresponding speed shell.
We refer \geqn{eq:standard spec} as the standard spectrum.
For the Maxwell distribution in \geqn{eq:Maxwell dist},
the integration can be performed analytically \cite{Derevianko:2016vpm, Foster:2017hbq},
\begin{equation}
\label{eq:standard spec for Max}
  S_{\phi} (f)
=
  \sqrt{\frac{2}{\pi}} \frac{1}{f_C\sigma v_{\odot}}
  \sinh \left( \frac{vv_{\odot}}{\sigma^2} \right)
  e^{- \frac {v^2 + v_{\odot}^2}{2 \sigma^2}} .
\end{equation}

\subsection{Doppler Effect}
\label{sec:Doppler effect}

The standard spectrum assumes a static detector in the SSB frame.
However, both heliocentric and ground-based detectors
move with respect to this frame and the spectrum in
\geqn{eq:standard spec} is modified by the Doppler effect.

It is straightforward to derive the spectrum measured by
a moving detector with relative velocity $\bf v'$.
This can be done by replacing $F(\mathbf{v})$ in
\geqn{eq:standard spec} with the corresponding
boosted distribution $F'(\mathbf{v}')$ in the detector frame.
By applying the Galilean transformation $\mathbf{v}' = \mathbf{v} - \mathbf{v}_{d}$,
where $\mathbf{v}_{d}$ is the detector velocity in the SSB frame,
the phase space distribution $F'(\mathbf{v}') = F(\mathbf{v}'+\mathbf{v}_{d})$
takes exactly the same functional form due to
the Liouville theorem.
Therefore, the standard spectrum in \geqn{eq:standard spec}
shifts with the Doppler effect as,
\begin{equation}
\label{eq:Doppler spectrum}
  S_{\phi}(f;\mathbf{v}_{d})
=
  \frac{v}{f_C} \int {\rm d}\hat{\mathbf{v}}  F(\mathbf{v}+\mathbf{v}_{d}).
\end{equation}
The spectrum exhibits time dependence through
the detector velocity ${\bf v}_d$, which will be
examined with further details in \gsec{sec:total effect}.

Besides the Earth motion around the Sun,
the Earth rotation can also contribute to the
Doppler effect for terrestrial detectors. The Earth surface
velocity $\mathbf{v}_s$ due to the Earth rotation
can be incorporated into the detector velocity as
${\bf v}_d = {\bf v}_\oplus \rightarrow {\bf v}_\oplus + {\bf v}_s$
in \geqn{eq:Doppler spectrum}
with ${\bf v}_\oplus$ being the Earth velocity.
However, even with the maximal Earth surface rotation speed
$|\mathbf{v}_s| \sim 0.5$\,km/s around the
equator, the induced correction
is only of $\mathcal{O}(0.1\%)$ at most and
hence can be safely neglected.

\gfig{fig:relative diff} shows the relative differences between the corresponding
spectra observed by a ground-based detector ($|\mathbf{v}_{d}| \approx 10^{-4}$)
at four representative
times of the year and the standard spectrum.
A correction of approximately
$5\% \sim 15\%$ arises at the low- and
high-frequency ends of the spectrum. 
The correction is most significant at the beginning
of June and December
when the Earth's velocity is aligned with the
DM wind \cite{Lee:2013wza}. On the other hand, the correction
is much smaller at the beginning of March and
September when the
Earth's velocity is approximately orthogonal
to the DM wind.
In June, the Earth's velocity is anti-parallel to the DM wind, so most DM particles have larger velocities in the detector frame, and the spectrum is blue-shifted.
Conversely, in December, the Earth's velocity is parallel to the DM wind, leading to smaller velocities in the detector frame and a red-shifted spectrum.

One can show that the Doppler effect redistributes
power across frequencies while leaving the total power
unchanged. In other words, the Doppler effect does
not alter the number density.
A direct consequence is that the correction
takes opposite signs at the low- and high-frequency
ends as clearly shown in the first panel of
\gfig{fig:relative diff}.
The transition point happens when the derivative,
\begin{equation}
\label{eq:S/vdot}
    \frac{\partial\ln S_{\phi}}{\partial \mathbf{v}_d}\Bigg|_{v}
    \simeq \left[
    \frac{v}{v_{\odot}\sigma^2}\coth\left(\frac{vv_{\odot}}{\sigma^2}\right)
    -\frac{1}{v_{\odot}^2}
    - \frac{1}{\sigma^2}\right]\mathbf{v}_{\odot},
\end{equation}
vanishes with
$v \simeq v_{\odot}(1 + \sigma^2/v_{\odot}^2) \approx 1.5v_{\odot}$
for the SHM.
At this point, the number of particles leaving
the speed shell equals the number entering
it, which leaves the total particle number
unchanged at the leading order.

It needs to be emphasized that
although the relative difference is not negligible at the high-frequency tail, the absolute
magnitude of the spectra gets
exponentially suppressed
and the data there are likely dominated by noise.
Therefore, we expect the Doppler effect to be
difficult to resolve in this regime.

\subsection{Gravitational Focusing}
\label{sec:GF}

Using \geqn{eq:phi hyp} and \geqn{def:psd},
it can be shown that the spectrum with GF is given by
\begin{equation}
\label{eq:static psd with GF}
    S_{\phi} (f;\mathbf{x}_d) = \frac{v}{f_C} \int {\rm d} \hat{\mathbf{v}}  F(\mathbf{v}) \left|\chi_{\mathbf{k}}(\mathbf{x}_d)\right|^2 ,
\end{equation}
where $\mathbf{x}_d$ denotes the detector location.
The magnitude of GF is controlled by
$\eta/z = GM_{\odot}/2r_dv^2\sin^2(\theta/2) \sim r_{\odot}/ 2 r_d$
with the scattering strength $\eta$ and parameter
$z$ defined below \geqn{def: chi},
While $r_{\odot}\equiv GM_{\odot}/v^2$ is the
influence radius of the Sun's gravitational field,
$r_d$ is the distance from the solar center.\footnote{For the discussions
in this and the subsequent sections, the following scaling relations and numerical values are useful. For $m=10^{-13}\,\text{eV}$, we have $\eta \propto m/v \sim 7.5\times10^{-4}/v$, $z \propto mv \sim 1.5\times10^5 v$, $\eta/z \propto v^{-2} \sim 4.9\times10^{-9}/v^2$, and $\eta z \propto m^2 \sim 1.1\times10^2$.}
Note that the ratio is independent of the ULDM mass $m$,
implying that the magnitude of the correction is also
largely mass independent. As the ULDM mass
$m$ decreases, the scattering strength $\eta \equiv GM_{\odot}m/v$
and hence the GF effect decrease. On the other hand,
its wavelength and consequently the characteristic length
scale of GF as illustrated in \gfig{fig:avg density contrast}
would increase. Relatively speaking, the detector
would move closer to the gravitational center or
equivalently the Sun to feel larger GF effect.
These two contributions cancel with each other to
leave the magnitude of GF almost unchanged.

The effect of the GF correction has been shown in the
middle panel of \gfig{fig:relative diff}. 
As expected, the relative difference is the largest on
March 1st when the detector is directly downstream
of the DM wind \cite{Lee:2013wza} and the smallest
on September 1st when it is upstream.
The relative differences on June 1st and December 1st
overlap due to the rotational symmetry of
GF around the bulk velocity $-{\bf v}_\odot$
that is shown as a black arrow in \gfig{fig:avg density contrast}.
A prominent feature is that for the four time
points, the GF corrects are all positive since
the GF enhances the density which is different
from the Doppler effect.
When $\tau_c(f-f_C) \approx 1$, the correction is less than $5\%$.
At higher frequencies with shorter wavelength,
the correction decreases monotonically because
the corresponding wave components have larger
asymptotic momenta and are therefore less scattered
by the Sun. As the frequency approaches $f_C$,
however, the relative difference increases rapidly
due to the strong scattering of low-momentum DM
particles by the solar gravitational potential.

For mass much larger than $10^{-14}$\,eV such
that the de Broglie wavelength $1/m v$ becomes
much smaller than the Solar-Earth
distance, the wave description can be approximated by
the particle picture \cite{Kim:2021yyo}.
Then we have $|\chi_{\mathbf{k}}(\mathbf{x})|^2 \simeq J(\mathbf{v},\mathbf{x})$
(see \gapp{app:semi-classical} for a heuristic proof),
\begin{equation}
  J(\mathbf{v},\mathbf{x})
\equiv
  \frac 1 2
\left[
  \left(1+\frac{4\eta}{z}\right)^{1/2}
+ \left(1+\frac{4\eta}{z}\right)^{-1/2}
\right],
\label{eq:Jacobian}
\end{equation}
which is the Jacobian matrix that relates the infinitesimal
velocity element at position $\mathbf{x}$ to the corresponding
one at the asymptotic infinity as derived in the context
of particle GF \cite{Sikivie:2002bj, Alenazi:2006wu}. 
Since $\eta/z$ is positive, the function
$J(\mathbf{v},\mathbf{x})$ is always greater
than unity, which explains the positiveness
of the GF correction
as shown in the middle panel of \gfig{fig:relative diff}. 
For $v \sim \sigma$, we have $\eta/z \sim 10^{-1}$ and $J(\mathbf{v},\mathbf{x}) - 1 \simeq 2 \eta^2 / z^2 \sim 10^{-2}$
which indicates a GF correction at the percentage level.
 
\begin{figure}
\centering
\includegraphics[width=0.9\linewidth]{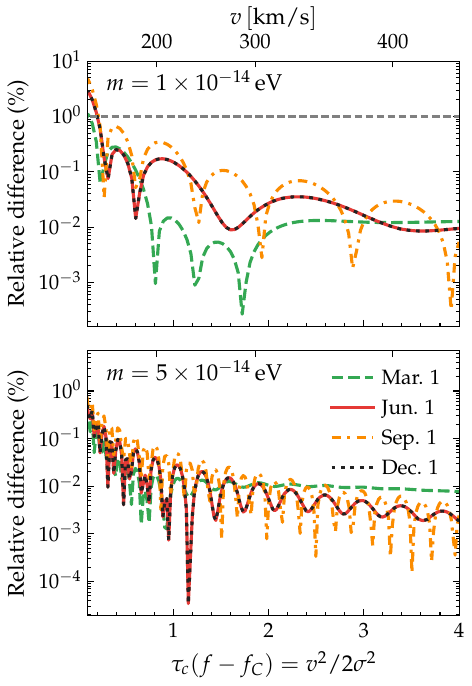}
\caption{Relative difference between the spectra
computed using the particle approximation and the
full result from Eq.\,(\ref{eq:static psd with GF})
on the first day of March (green dashed), 
June (red solid), September (orange dash-dotted),
and December (black dotted).
The two panels show the cases of $m=10^{-14}\,\text{eV}$ (Upper) and $m=5\times10^{-14}\,\text{eV}$ (Lower), respectively.
The dashed gray line in the upper panel denotes a relative difference of $1\%$.
}
\label{fig:jacobian_vs_chi}
\end{figure}

\gfig{fig:jacobian_vs_chi} shows the relative difference between the spectra obtained by replacing $|\chi_{\mathbf{k}}(\mathbf{x})|^2$ with $J(\mathbf{v},\mathbf{x})$ in \geqn{eq:static psd with GF} and the original spectra given by \geqn{eq:static psd with GF}.
We show the cases of $m=10^{-14}\,\text{eV}$ and
$m=5\times10^{-14}\,\text{eV}$ that corresponds to
$m v r \approx 4$ and $20$, respectively.
Even for $m v r \sim \mathcal{O}(1)$, the difference remains well below the typical magnitude of GF correction over nearly the entire frequency band.

\subsection{Total Effect}
\label{sec:total effect}

Now we are in a position to discuss the spectrum including
both the Doppler and GF effects. Using the Lorentz
transformation to combine \geqn{eq:Doppler spectrum}
with \geqn{eq:static psd with GF},
the spectrum including both effects is given by
\begin{equation}
\label{eq:full spec}
  S_{\phi} (f)
=
  \frac v {f_C}
  \int {\rm d} \hat{\mathbf{v}} F(\mathbf{v}+\mathbf{v}_{d}) \left|\chi_{\mathbf{k}+\mathbf{k}_{d}}(\mathbf{x}_d)\right|^2,
\end{equation}
when neglecting the variations in the detector velocity
and $\chi_{\mathbf{k}}(\mathbf{x})$ along its orbit.
As shown in the right panel of \gfig{fig:relative diff},
including both effects breaks the degeneracy between March
and September for the Doppler effect as well as the
degeneracy between June and December for GF. 
At the high-frequency tail, the correction is dominated
by the Doppler effect as GF becomes less significant
for particles with larger momenta. Towards the low-frequency
end, GF becomes increasingly important and dominates
the total correction.

A natural question is when the corrected
spectrum would deviate the most from the standard case. 
In the signal-dominated scenario, the data likelihood
is given by
\begin{equation}
\hspace{-3mm}
- 2 \ln\mathcal{L}  
\equiv
  \frac {N_{\text{cyc}} }{f_C}
  \sum_{i=1}^n \Delta f
\left[
  \frac {\hat{S}_{d}(f_i) - S_{\phi}(f_i;\mathbf{v}_{d},\mathbf{x}_d)}
        {S_{\phi}(f_i;\mathbf{v}_{d},\mathbf{x}_d)}
\right]^2.
\label{eq:s-dominated likelihood}
\end{equation}
Note that we have omitted a $\ln S_{\phi}(f_i)$ term that also depends
on the spectral model, since its variation as the
parameters change is negligible for $N\gg1$, and we absorb
it as a constant term. The above formula suggests a
{\it mismatch factor} $\gamma$ to describe
the difference between the corrected spectrum and its
standard counterpart,
\begin{equation}
  \gamma
\equiv
  \tau_c
  \int_{f^{\text{low}}_{1/2}}^{f^{\text{high}}_{1/2}} {\rm d}f
\left[ \frac{S_{\phi}(f;\mathbf{v}_{d},\mathbf{x}_d) - S_{\phi}(f)}{S_{\phi}(f)} \right]^2 ,
\label{eq:mismatch}
\end{equation}
where we integrate over the frequency range from $f^{\text{low}}_{1/2} \approx f_C + 0.13\tau_c^{-1}$ to $f^{\text{high}}_{1/2} \approx f_C + 3.48\tau_c^{-1}$ specified by $S_{\phi}(f) = \max[S_{\phi}]/2$.

\begin{figure}
\centering
\includegraphics[width=0.9\linewidth]{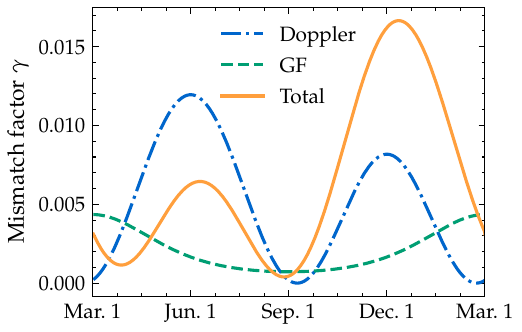}
\caption{Evolution of the {\it mismatch factor}
$\gamma$ over the course of a year for three
scenarios: including only the Doppler effect
(blue dash-dotted), including only the GF
(green dashed), and the total result including
both effects (orange solid).
}
\label{fig:mismatch}
\end{figure}

\gfig{fig:mismatch} shows the evolution of $\gamma$ over the course of a year for three scenarios: including only the Doppler effect, including only the GF, and including both.
When including only the Doppler effect, the maximum mismatch
appears around June 1st and December 1st which is consistent
with the discussions in \gsec{sec:Doppler effect}.
If only the GF effect is included, the largest deviation appears
around March 1st when the Earth is downstream of the DM wind.
When both effects are included, while $\gamma$ is predominantly shaped by the Doppler correction,
GF still plays a significant role. 
Specifically, GF suppresses $\gamma$ around June but
enhances it around December due to destructive and
constructive superpositions. 
This can be seen by comparing the first two panels
of \gfig{fig:relative diff}. Below 300\,km/s where
most of the phase space resides, the Dopper effect
is negative in June while the GF focusing is positive.
On the other hand, both have positive values in December.
Consequently, the maximum moves from June, when only the Doppler correction is considered, to December when both effects are included.
The GF effect also shifts the peak of $\gamma$ toward March for both peaks.
Consequently, the time at which the total correction is most
significant is not December 1st but a few days later.
Similar thing happens for the minor peak around June 1st.
This is analogous to the annual modulation of the signal in the detection of particle DM \cite{Lee:2013wza}.
The correction in September is not very pronounced, as the detector is upstream of the DM wind and its velocity is also orthogonal to the wind such that neither effect is significant.

\subsection{Spectral Template}
\label{sec:template}

The spectrum in \geqn{eq:full spec} is derived assuming
a constant velocity and neglecting the spatial variation of GF along the Earth orbit.
However, the above assumptions can only hold approximately
over certain timescales as detectors orbit around the Sun.
While the spectrum can in principle be computed numerically
using the time-domain simulations described below,
the computational cost and runtime make it
impractical for data analysis. 
We therefore seek to construct an
approximate spectrum that is computationally efficient while
retaining sufficient accuracy to capture these effects.
Since both effects modify the spectrum at the percentage level,
we require the approximation to achieve a comparable accuracy.

There are three timescales involved in this problem.
{\bf (1)} The first is the {\it coherence time},
\begin{equation}
  \tau_c
\equiv
  \frac {2 \pi}{m\sigma^2}
\approx
  42.6\,\text{h} \left(\frac{10^{-13}\,\text{eV}}{m}\right)\left(\frac{5.2\times10^{-4}}{\sigma}\right)^2 .
\end{equation}
Since the spectrum has a width of $\sim\mathcal{O}(\tau_c^{-1})$ in the Fourier domain, the data duration $T$ must be longer than $\tau_c$ to resolve the spectrum.
{\bf (2)} The second time scale is associated with the Earth's orbital motion.
We define $T_{\text{orb}}$ as the time scale over which the deviation of the Earth's orbital motion from the uniform linear motion introduces a correction smaller than a prescribed threshold. Over a duration of $T_{\text{orb}}$, the change in the Earth's velocity is $|\Delta\mathbf{v}_d| / v_d \simeq \omega_d T_{\text{orb}}$ where $\omega_d$ is the Earth's orbital angular velocity. 
Using \geqn{eq:S/vdot}, the variation in spectrum due to this can be estimated as
\begin{equation}
    \Delta\ln S_{\phi}
    \simeq
    \frac{\partial\ln S_{\phi}}{\partial \mathbf{v}_d} \Delta\mathbf{v}_d
    \sim \frac{\mathbf{v}_{\odot} \cdot \Delta\mathbf{v}_d}{v_{\odot}^2},
\end{equation}
where the right-hand side applies at the low-frequency end.
For $T_{\text{orb}} \lesssim 5.8\,\text{days}$, $|\Delta\mathbf{v}_d|/v_{\odot} \lesssim 1\%$
and the resulting correction remains below $1\%$.
{\bf (3)} The third time scale comes from the GF effect.
As the Earth moves along its orbit, it explores regions with different GF strengths. Because the Earth follows a nearly circular orbit around the Sun, the variation in the GF effect along the orbit is primarily caused by the change in the angle between the Earth's position vector and the DM wind direction.
The angular dependence of the GF effect can be quantified by 
\begin{equation}
    \frac{d\ln (J-1)}{d\theta} = -\frac{1}{2}\left(1 + \frac{1}{\sqrt{1+4\eta/z}}\right)^2
    \cot\left(\frac{\theta}{2}\right) ,
\end{equation}
which gives the relative size of variation with respect to the GF effect that
is proportional to $J - 1$.
The variation in the spectrum over an interval $T_{\text{GF}}$ can then be estimated as 
\begin{equation}
    \Delta\ln S_{\phi}
    \sim \Delta\ln J = \frac{J - 1}{J} \frac{d\ln (J-1)}{d\theta} \omega_d T_{\text{GF}} ,
\end{equation}
where $\omega_d T_{\text{GF}}$
gives the typical angular movement over the interval.
From \geqn{eq:Jacobian}, we find $(J-1)/J \simeq 1$ for
$\eta/z \gg 1$ at the low-frequency end of the spectrum
while $(J-1)/J \simeq J-1 \ll 1$ for $\eta/z \ll 1$
at the high-frequency end.
Given that $d\ln (J-1)/d\theta\sim \mathcal{O}(1)$,
we have $\Delta\ln S_{\phi} \sim \omega_d T_{\text{GF}}$
at the low-frequency end while the higher-frequency
has $\Delta\ln S_{\phi} \sim (J-1)\omega_d T_{\text{GF}}$
which is suppressed by $J-1 \ll 1$.
In other words, the major variation comes from the
low-frequency region. Requiring the correction at the
low-frequency end to remain below $1\%$ gives
$T_{\text{GF}} \lesssim 0.6\,\text{days}$, which
is more restrictive than the constraint on $T_{\text{orb}}$.
However, the
low-frequency region that provides the large variation
is confined within a narrow frequency range very close
to the Compton frequency $f_C$. For example, even at
$\tau_c(f-f_C) = 0.5$, we have $\eta/z \sim 1.8\times10^{-2} \ll 1$ and hence does not contribute much.
Consequently, only a negligible fraction of data points will fall within this region and they contribute little to the overall signal statistic.

Therefore, within a duration of $\tau_c < T \lesssim T_{\text{orb}}$, the spectral template including both effects can then be given by
\begin{equation}
\label{eq:spec template}
  S_{\phi} (f;\mathbf{v}_{d},\mathbf{x}_d)
\simeq
  \frac{v}{f_C} \int {\rm d}\hat{\mathbf{v}} 
  F(\mathbf{v}+\mathbf{v}_{d})
  J(\mathbf{v}+\mathbf{v}_{d},\mathbf{x}_d).
\end{equation}
where we have replaced $|\chi_{\mathbf{k}}(\mathbf{x})|^2$
with $J(\mathbf{v},\mathbf{x})$ for $m \gtrsim 5\times10^{-14}\,\text{eV}$, which introduces errors smaller than $1\%$. 
This improves the efficiency of integration considerably, since computing $\chi_{\mathbf{k}}(\mathbf{x})$ becomes increasingly expensive as the mass increases.
By contrast, the Jacobian matrix involves only elementary functions and can be evaluated efficiently.
Another advantage of the template in \geqn{eq:spec template} is that it has a universal shape that is independent of the ULDM mass. 
This could be exploited in data analysis, since the integral in \geqn{eq:spec template} is now mass independent and can be precomputed for a given velocity distribution.

\begin{figure}
\centering
\includegraphics[width=0.9\linewidth]{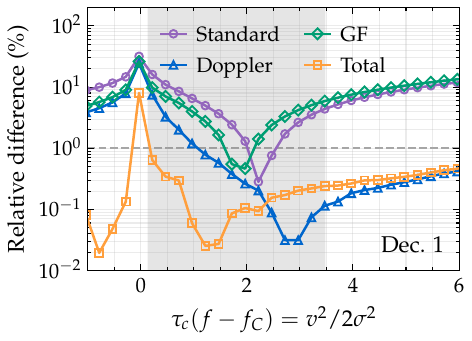}
\caption{Relative differences between the numerical
spectrum and four analytical spectra: standard
(purple circles), Doppler-corrected (blue triangles),
GF-corrected (green diamonds), and the total template
with both effects included (orange squares). 
The gray shaded region indicates the frequency
range where the spectral amplitudes exceed
$\max[S_{\phi}]/2$.}
\label{fig:residual}
\end{figure}

We numerically compute the spectrum for $m=5\times10^{-13}\,\text{eV}$ over $T=4\tau_c \approx1.4\,\text{days}$ around December 1st.
\gfig{fig:residual} shows the relative differences between the numerical spectrum and the spectra given in Eq.\,(\ref{eq:standard spec}), (\ref{eq:Doppler spectrum}), (\ref{eq:static psd with GF}), and (\ref{eq:spec template}).
We find that the relative difference between the numerical spectrum and the template in \geqn{eq:spec template}
with both the Dopper and GF effects included
remains below $1\%$ with the only exception of a signle
point at the Compton frequency $f_C$.
If only the Dopper or GF effect is included, the
deviation would become much larger across the whole
frequency range. The situation would become much
worse if neither effect is taken into account.
The results presented here provide a first demonstration of the effectiveness of our spectral template.
Its performance is further assessed below by performing Bayesian inference on the simulated data.

\section{Mock data and Bayesian inference}
\label{sec:MD & Bayesian}

To illustrate the impact of these effects, we perform Bayesian inference on mock data sets using the developed spectral templates. 
Before delving into the numerical results, it will be helpful to derive some analytical results that provide insight into their interpretation.

We denote the data time series by $d(t)$ which consists of both
signal $s(t)$ and detector noise $n(t)$. 
The PSD of data is given by \geqn{def:psd} with $\tilde{\phi}(f)$
replaced by the Fourier transform $\tilde{d}(f)$ of the data.
When the signal and noise are uncorrelated, the data
PSD can be decomposed as $S_{d}(f) \equiv S_{\phi}(f) + S_{n}(f)$
where $S_{n}(f)$ is the noise PSD.

\subsection{Analytical Estimation of Signal Significance}
\label{sec:snr criterion}

The definition in \geqn{def:psd} corresponds to an ensemble average in the limit of infinitely many realizations, or equivalently, an infinitely long observation time.
In practice, however, experiments have a finite duration,
and the ensemble average is replaced by a finite average, 
\begin{equation}
\label{def:psd measured}
  \hat{S}_d(f)
\equiv
  \frac{2}{N}\sum_{a=1}^{N} \frac{|\tilde{d}_a(f)|^2}{T} ,
\end{equation}
where the data of total duration $T_{\text{obs}}$
are divided into $N$ segments, each with a duration
$T$. The Fourier transform $\tilde{d}_a(f)$ of data
is for the $a$-th segment.
Since the spectrum has a width $\sim1/\tau_c$ in the Fourier space, the segment duration $T$ needs to be longer than $\tau_c$ to resolve the spectral profile.
The measured PSD $\hat{S}_d$ exhibits
fluctuations and deviates from the expectation value $S_{d}$.
The central limit theorem \cite{Jaynes_2003, Gregory_2005} implies that $\hat{S}_d$
follows a Gaussian distribution with mean $S_{d}$ and variance $S_{d}^2 / N$.
Therefore, in the signal-dominated regime ($S_{d} \simeq S_{\phi}$), averaging over $N$ segments suppresses the fluctuations to $S_{\phi}/\sqrt{N}$.
The correction become relevant when the difference between the corrected and standard spectra is comparable to or larger than the fluctuations, i.e.,
$S_{\phi}(f;\mathbf{v}_d,\mathbf{x}_d)-S_{\phi}(f) \gtrsim S_{\phi}(f)/\sqrt{N}$, or equivalently
$\sqrt{N} \gtrsim \epsilon^{-1}$ where $\epsilon$
denotes the relative difference between the corrected
and standard spectra.
For a typical correction of $\epsilon \sim 10\%$, this requires $N\gtrsim10^2$.

The above discussion of the number of data segments can be recast in terms of the signal-to-noise ratio (SNR) \cite{Thrane:2013oya},
\begin{equation}
\label{def: SNR}
  \text{SNR}
\equiv
  \sqrt{T_{\text{obs}}
  \int_{0}^{\infty} {\rm d} f
  \left( \frac{S_{\phi}(f)}{S_n(f)} \right)^2
  },
\end{equation}
which is more closely related to detection.
As the ULDM spectrum is narrow band, the noise
PSD can be approximated as white over the band.
Or equivalently, the noise becomes a $S_n$.
Then \geqn{def: SNR} to be written as
\begin{equation}
  \text{SNR}
  =
  \sqrt{N T f_C \Upsilon}
  \frac{S_{\phi,\max}}{S_n}
  ,
\label{eq:SNR 1}
\end{equation}
where $S_{\phi,\max} \equiv \max [S_{\phi}(f)]$
is the spectral maximum and the form factor,
\begin{equation}
  \Upsilon
\equiv
  \frac 1 {f_C}
  \int_{f_C}^{\infty} df
  \left( \frac{S_{\phi}(f)}{S_{\phi,\max}} \right)^2,
\end{equation}
carries the information of spectral shape.
Note that there is no actual contribution
bellow the Compton frequency $f_C$.
For the velocity distribution in \geqn{eq:Maxwell dist}
and the SHM parameters \cite{Drukier:1986tm,Freese:1987wu,Bozorgnia:2016ogo},
one may first evaluate a concrete value
$\Upsilon \approx 6.8 \times 10^{-7}$.
Putting the requirement $\sqrt{N} \gtrsim \epsilon^{-1}$ into \geqn{eq:SNR 1} and taking $T=\tau_c$ and $f_C\tau_c = 1/\sigma^2$, we obtain
\begin{equation}
\label{eq:snr criteria}
  \text{SNR}
\gtrsim
  159 
  \left( \frac{10\%}{\epsilon} \right)
  \left( \frac{S_{\phi,\max}/S_n}{10} \right) .
\end{equation}
To put this SNR into context, we consider gravitational-wave interferometer experiments as a means of probing ULDM.
Although designed for gravitational-wave detection,
the ground-based laser interferometers such as LIGO
are also sensitive to ULDM through various interactions
between ULDM and their components. For example,
the dark photon field can induce oscillations of
the test masses as in \geqn{eq:dp_acc}.
Existing constraints have been obtained using the
LIGO data \cite{Guo:2019ker,LIGOScientific:2021ffg}.
If an ULDM signal lies just below the current sensitivity, it would have an SNR of $\mathcal{O}(10^2\sim10^4)$ in the third-generation interferometers such as the Einstein Telescope \cite{ET:2019dnz} and Cosmic Explorer \cite{Evans:2021gyd}, whose noise performance is expected to improve by several orders of magnitude than LIGO.

For completeness, we also briefly discuss the noise-dominated regime.
In this case, the fluctuations of $\hat{S}_d$ are dominated by the detector noise and are suppressed to $S_n/\sqrt{N}$ after averaging over $N$ segments.
As in the signal-dominated case, the correction can only be resolved when the difference between the corrected and standard spectra is comparable to or larger than the measurement uncertainty, which in this case is given by $S_n/\sqrt{N}$, i.e., $S_{\phi}(f;\mathbf{v}_d,\mathbf{x}_d)-S_{\phi}(f) \gtrsim S_n/\sqrt{N}$, or equivalently $\sqrt{N} \gtrsim \epsilon^{-1}S_n/S_{\phi}$.
Substituting this condition into \geqn{eq:SNR 1}, we obtain
\begin{equation} \label{eq:snr criteria noise-dominated}
    \text{SNR} \gtrsim 16\left(\frac{10\%}{\epsilon}\right) .
\end{equation}

\subsection{Mock Data}
\label{sec:mock data}
In this section, we describe how we generate the simulated data.
We begin with a time domain simulation of the field using 
\begin{equation} \label{eq:phi along orbit}
    \phi(x) = \sum_{\mathbf{v}}  \sqrt{\Delta_{\mathbf{v}}} \beta_{\mathbf{v}}
    e^{-i\left(\omega t - \mathbf{k}\cdot\mathbf{x}(t) + \varphi_{\mathbf{v}}\right)}
    \chi_{\mathbf{k}}(\mathbf{x}(t)) ,
\end{equation}
where $\mathbf{x}(t)$ is the detector's trajectory in the SSB coordinates.
The discrete sum over velocity above is performed by dividing the velocity space into uniformly spaced shells, with each shell further partitioned into equal-area elements using the \href{https://healpix.sourceforge.net}{\texttt{healpy}} package.\footnote{\href{ https://healpix.sourceforge.net}{https://healpix.sourceforge.net}}
According to the dispersion relation $df/f_C = vdv$,
this sampling scheme provides a smaller $d f$
for smaller velocity $v$ and hence a larger number
of sample points as the frequency approaches $f_C$.
This feature is desirable because the spectrum varies rapidly near $f_C$ and a finer frequency grid in this region helps reduce numerical errors.
We adopt $200$ shells and set $\text{nside}=16$ in \texttt{healpy}, which corresponds to $6.1\times10^5$ sampling points in velocity space.\footnote{The specific choices for the number of shells and $\text{nside}$ are guided by numerical convergence tests, in which we vary parameters and verify that the computed spectrum has converged for the chosen values.} 
The time domain series is then used to compute the spectrum according to \geqn{def:psd measured}.
Note that, as the signal is generated in the SSB coordinates, the time axis is rescaled using the Lorentz transformation before computing the PSD in the detector frame.

\begin{figure}
\centering
\includegraphics[width=0.9\linewidth]{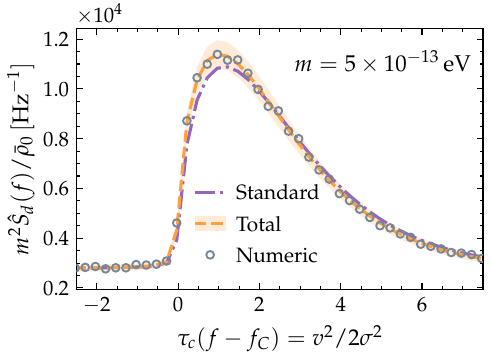}
\caption{The numerical PSD (gray circle)
computed according to Eq.\,(\ref{def:psd measured})
using $3600$ segments, each with a duration of $4\tau_c$. 
Also shown are the PSDs (signal plus noise) obtained
from the standard spectrum (purple dash-dotted) and
the total spectral template (orange dashed).
The light orange band indicates the $3\sigma$
uncertainty associated with the template.}
\label{fig:data psd}
\end{figure}

We simulate signal for a ULDM of $m=5\times10^{-13}\,\text{eV}$ around December 1st when the mismatch is the largest. 
Each signal segment spans a duration of $T=4\tau_c\approx1.4$\,days and is sampled at a rate of $2.1f_C$, resulting in approximately $3.1\times10^7$ data points.
Note that the choice of mass is for illustration purposes and the following results should be also valid for other masses in the semiclassical regime as the spectrum exhibits a universal shape.
The signal then mixes with a white noise with an amplitude of $S_n = S_{\phi,\max}/3$. 
The data PSD, computed using $3600$ segments, is shown in \gfig{fig:data psd}.
The corresponding SNR calculated using \geqn{eq:SNR 1} is $573$. 
For comparison, we also show the theoretical PSDs obtained from the standard spectrum in \geqn{eq:standard spec} and the template in \geqn{eq:spec template}.
The light orange band indicates the $3\sigma$ region associated with the template. As shown in the figure, the template agrees well with the PSD of the simulated data, whereas the standard spectrum exhibits clear deviations.
Note that the finite duration of each segment causes spectral leakage, which smooths the sharp cutoff implied by the dispersion relation into a more gradual transition in the data PSD near $f_C$.

We note that the coherence time becomes longer for lighter ULDM and the detector motion over a coherence time can no
longer be approximated as constant as discussed in Sec.~\ref{sec:template}. Moreover,
the wave behavior also becomes significant in
this regime. Although obtaining an analytical
expression for the spectrum is difficult in this
case, the signal can still be simulated using
\geqn{eq:phi along orbit}, and the spectrum can
then be computed numerically.
However, detecting the corrections may not be
easy for lighter ULDM. As discussed in
\gsec{sec:snr criterion}, observing these
corrections requires a sufficient number of
independent realizations of the signal.
As the mass decreases, the coherence time grows. 
For $m \sim  10^{-15}\,\text{eV}$, $\tau_c \sim 50\,\text{days}$,
so $100\tau_c \sim 14\,\text{yr}$, which exceeds
the lifetime of most experiments.

\subsection{Bayesian Inference}
\label{sec:Bayesian infer}

Since the solar velocity $\mathbf{v}_{\odot}$
appears in \geqn{eq:spec template},
we can extract its directional information or more
precisely the component parallel to the ecliptic plane.
This should be contrasted with the standard spectrum,
where only the speed $v_{\odot} = |\mathbf{v}_{\odot}|$
appears in \geqn{eq:standard spec for Max} since the
directional information is completely lost after angular integration.
While it worth exploring how well the directional
information can be extracted,
here we restrict ourselves to the four parameters
that fully determine the standard spectrum: the
ULDM mass $m$, the solar speed $v_{\odot}$, the
velocity dispersion $\sigma$, and the local DM
energy density $\bar\rho_0$.

\begin{figure*}[t]
\centering
\includegraphics[width=0.49\linewidth]{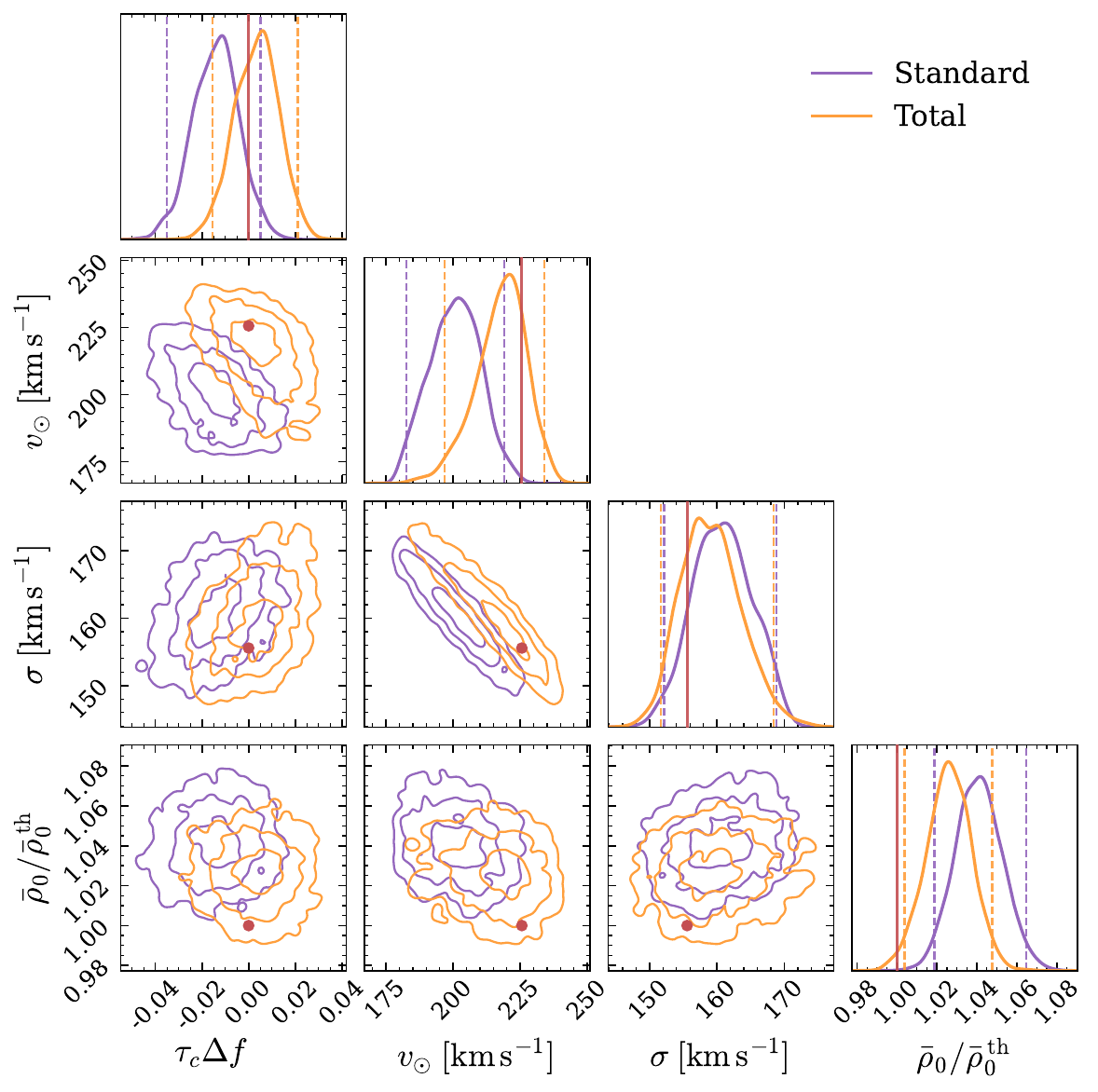}
\includegraphics[width=0.49\linewidth]{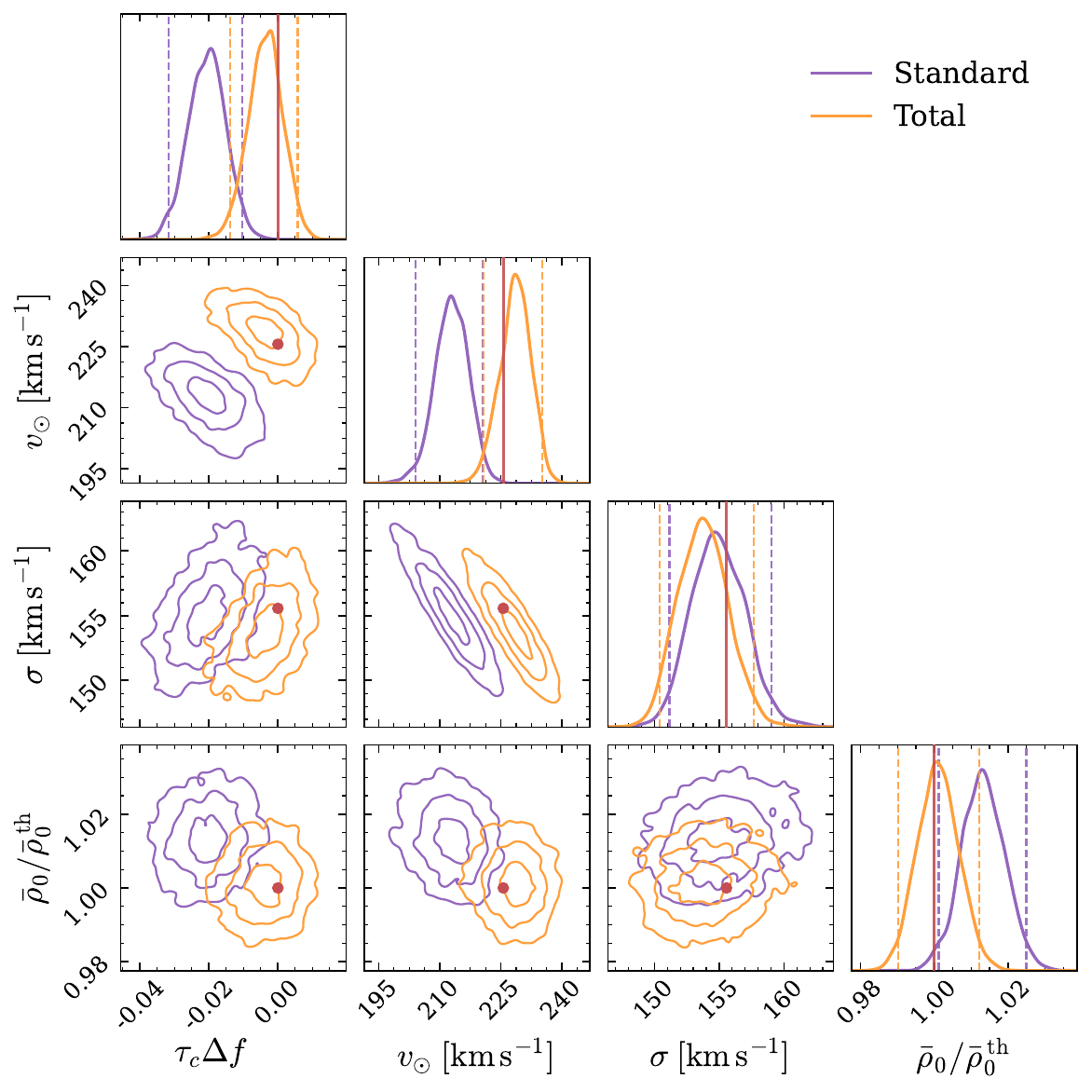}
\caption{Posteriors of the parameters
inferred using the standard spectrum (purple) and the template incorporating both effects (orange).
In the diagonal panels, vertical dashed lines in the corresponding colors mark the bounds of the $95\%$ credible intervals. Red dots and solid lines indicate the input values.
The left and right panels show results for $900$ and $3600$ segments, corresponding to SNRs of $287$ and $573$, respectively.
} 
\label{fig:posterior normal set}
\end{figure*}

To make comparison with the mock data,
our data likelihood is given by
\begin{equation}
\label{eq:full likelihood}
\hspace{-3mm}
    -2\ln\mathcal{L}(\hat{S}_d|\boldsymbol{\lambda})
    = \sum_{i=1}^n N\frac{(\hat{S}_{d,i} - S_{d,i})^2}{S_{d,i}^2}
    + \ln\left(\frac{2\pi S_{d,i}^2}{N}\right) ,
\end{equation}
where $\boldsymbol{\lambda} \equiv [m,v_{\odot},\sigma,\bar\rho_0]$
collects signal parameters and $S_{d,i} \equiv S_{d}(f_i)$
denotes the projected value at the $i$th frequency point.
We compare the performance of two spectral models:
the standard spectrum in \geqn{eq:standard spec}
and the corrected template in \geqn{eq:spec template}.
The theoretical data PSD entering \geqn{eq:full likelihood}
is given by $S_d \equiv S_{\phi} + S_n$ where
$S_{\phi}$ is evaluated using the corresponding
spectral model. We take the mass prior to be uniform
over $\tau_c\Delta f \in [-5, 5]$ where
$\Delta f \equiv f_C - f_C^{\text{th}}$ and
$f_C^{\text{th}}$ denotes the true value.
For the remaining parameters
$\theta \in \{v_{\odot}, \sigma, \bar\rho_0\}$,
we use log-uniform priors over
$\theta/\theta^{\text{th}} \in [0.8, 1.2]$
where $\theta^{\text{th}}$ denotes the true
value of each parameter. 
The posteriors are computed using {\tt dynesty} \cite{Speagle:2019ivv, sergey_koposov_2025_17268284, Skilling:2004pqw, Skilling:2006gxv, Feroz:2008xx}
and are shown in \gfig{fig:posterior normal set}.

The left panel of \gfig{fig:posterior normal set}
shows the results for $900$ segments, which has an SNR of $287$.
At this SNR, the standard spectrum recovers the
mass and velocity dispersion within the statistical
uncertainty but yields a biased estimate of the
solar speed $v_{\odot}$. For comparison, the corrected
template successfully recovers this parameter with
the best fit points very close to the corresponding
true values. The deviation in the solar speed is expected because the
detector moves in the direction of DM wind on December 1,
which leads to a smaller wind speed in the detector frame.
Notably, the posterior obtained for all four parameters
using the corrected template lie closer to the true values than
those obtained with the standard spectrum.
The true density lies outside the $95\%$ credible intervals for both models, which we attribute to statistical fluctuations in this particular realization of data.

The right panel shows the results for $3600$ segments.
With four times more segments, the statistical
fluctuations are suppressed to $\sim 2\%$ and
the effects of GF become more noticeable.
In this case, the standard spectrum can recover
only the velocity dispersion within the statistical
uncertainty, while the corrected template recovers
all four parameters. These results validate our
spectral template in \geqn{eq:spec template} with
both the Doppler and GF effects incorporated.

\section{Conclusion and discussions}
\label{sec:conclu_disc}

The gravitational focusing effect of wavy DM in
the solar gravitational potential is essentially
quantum interference at the solar system
scale. It can alter the ULDM density
around the Earth orbit at the percentage level
after averaging over the DM velocity distribution.
Such percentage effect is actually a reflection
of the DM wind in the solar frame. Since the signal
power recorded by a detector is approximately proportional
to the ULDM density for a linear coupling, such
variation in the density can affect the signal power
with comparable size.
Besides, the same DM wind velocity would lead to
Dopper effect of around the same size.

We compute the spectrum accounting for both effects
and find corrections at the percentage level over most
of the spectrum, where the Doppler effect dominates
as shown in \gfig{fig:relative diff}. 
The corrections become larger near the ULDM Compton
frequency $f_C \equiv m/2\pi$, where GF becomes
significant for low-speed DM particles.
We introduce a mismatch factor to quantify the
difference between the corrected and uncorrected
spectra in \gfig{fig:mismatch} and find that this factor varies annually,
reaching its maximum around December. This annual pattern is a smoking gun
for wavy DM and could help distinguish the signal from noise.

To assess the detectability of these corrections,
we first provide an analytical criterion based on
the signal-to-noise ratio. Then, we generate
simulated data containing both the ULDM signal
and detector noise followed by Bayesian
inference with the computationally efficient
spectral template developed in this work. 
Our results in \gfig{fig:posterior normal set}
show that accounting for the GF
and Doppler corrections in the signal-dominated regime
is essential to reliably recover the underlying
DM properties (such as the mass and DM wind velocity) once the integration time exceeds
$\mathcal{O}(10^2)$ coherence times.

\section*{acknowledgment}
YH thanks Hyungjin Kim for helpful discussion.
We acknowledge the use of NumPy \cite{harris2020array}, SciPy \cite{2020SciPy-NMeth}, mpmath \cite{mpmath}, 
healpy \cite{Zonca2019, 2005ApJ...622..759G}, dynesty \cite{Speagle:2019ivv, sergey_koposov_2025_17268284, Skilling:2004pqw, Skilling:2006gxv, Feroz:2008xx} in our code. All plots in this work are produced using Matplotlib \cite{Hunter:2007}.
SFG is supported by the National Natural Science
Foundation of China (12425506 and 12375101).
This work is also supported by State Key Laboratory of Dark Matter Physics.
SFG is also an affiliate member of Kavli IPMU, University of Tokyo.

\appendix
\section{Semi-classical limit} \label{app:semi-classical}
The quantum mechanical analog of the classical phase space distribution is given by the following quasiprobability distribution \cite{Wigner:1932eb,Kim:2021yyo}
\begin{align}
  F_W
& \equiv
  F_W(\mathbf{x}, \mathbf{p})
\\
& =
  \int d^3 {\bf y} d^3 {\bf v}
  e^{i \mathbf{p} \cdot \mathbf{y}}
  F(\mathbf{v})
  \psi_{\mathbf{v}}^* \left( \mathbf{x} + \frac{\mathbf{y}}{2} \right)
  \psi_{\mathbf{v}} \left( \mathbf{x} - \frac{\mathbf{y}}{2} \right),
\nonumber
\end{align}
where $\psi_{\mathbf{v}}(\mathbf{x})$ is the wave function.
Using the Schrödinger equation Eq.~(\ref{eq:Schrödinger equation}), it can be shown that $F_W(\mathbf{x}, \mathbf{p})$ follows
\begin{equation}
\begin{split}
    &\frac{\partial F_W}{\partial t}+\mathbf{v} \cdot \frac{\partial F_W}{\partial \mathbf{x}} \\
    &+ i\left[V\left(\mathbf{x}+\frac{i}{2} \nabla_{\mathbf{p}}\right) -V\left(\mathbf{x}-\frac{i}{2} \nabla_{\mathbf{p}}\right)\right] F_W=0 ,
\end{split}
\end{equation}
where $V\left(\mathbf{x}\right)$ is the potential.
When $|\nabla \ln V| \ll |\mathbf{p}|$, it reduces to the Boltzmann equation
\begin{equation} \label{eq:Boltzmann equation}
\frac{\partial F_W}{\partial t}+\mathbf{v} \cdot \frac{\partial F_W}{\partial \mathbf{x}} 
+ \mathbf{a} \cdot \frac{\partial F_W}{\partial \mathbf{p}} = 0 ,
\end{equation}
where $\mathbf{a} = -\nabla V(\mathbf{x})$.
For a stationary solution $\partial F_W/\partial t = 0$, Eq.~(\ref{eq:Boltzmann equation}) implies that $F_W(\mathbf{x}, \mathbf{p})$ is conserved along particle trajectories:
\begin{equation}
    F_W(\mathbf{x}_1, \mathbf{p}_1) = F_W(\mathbf{x}_2, \mathbf{p}_2) ,
\end{equation}
where $(\mathbf{x}_1, \mathbf{p}_1)$ and $(\mathbf{x}_2, \mathbf{p}_2)$ are two points connected by a trajectory.
In the following, we take $(\mathbf{x}_1, \mathbf{p}_1)$ for the Earth and $(\mathbf{x}_2, \mathbf{p}_2)$ for the asymptotic infinity.
At the asymptotic infinity, we have
\begin{align}
    &F_W(\mathbf{x}_2, \mathbf{p}_2) \\
    &= \int d^3 y e^{i \mathbf{p}_2 \cdot \mathbf{y}} \int d^3 v F(\mathbf{v}) \psi_{\mathbf{v}}^*(\mathbf{x}_2+\mathbf{y} / 2) \psi_{\mathbf{v}}(\mathbf{x}_2-\mathbf{y} / 2) \nonumber \\
    &= \int d^3 y e^{i (\mathbf{p}_2 - \mathbf{p}) \cdot \mathbf{y}} \int d^3 v F(\mathbf{v}) = \int d^3 v F(\mathbf{v}) \delta^{(3)}(\mathbf{p}_2 - \mathbf{p}) , \nonumber
\end{align}
where we use $\psi_{\mathbf{v}}(\mathbf{x}) = e^{i\mathbf{p}\cdot\mathbf{x}}$.
On the other hand, 
\begin{align}
    &F_W(\mathbf{x}_1, \mathbf{p}_1) \nonumber\\
    &= \int d^3 y e^{i \mathbf{p}_1 \cdot \mathbf{y}} \int d^3 v F(\mathbf{v})
    \psi_{\mathbf{v}}^*(\mathbf{x}_1+\mathbf{y} / 2) \psi_{\mathbf{v}}(\mathbf{x}_1-\mathbf{y} / 2) \nonumber\\
    &\simeq
    \int d^3 y e^{i \mathbf{p}_1 \cdot \mathbf{y}} \int d^3 v F(\mathbf{v}) 
    |\chi_{\mathbf{k}}(\mathbf{x}_1)|^2 e^{-i\mathbf{p}\cdot\mathbf{y}}  \nonumber\\
    &=  \int d^3 v F(\mathbf{v})|\chi_{\mathbf{k}}(\mathbf{x}_1)|^2 \delta^{(3)}(\mathbf{p}_1 - \mathbf{p}) ,
\end{align}
where we use $\psi_{\mathbf{v}}(\mathbf{x}) = \chi_{\mathbf{k}}(\mathbf{x}) e^{i\mathbf{p}\cdot\mathbf{x}}$ and take $\chi_{\mathbf{k}}(\mathbf{x} - \mathbf{y}/2) \simeq \chi_{\mathbf{k}}(\mathbf{x})$, since the contribution to the integration mainly comes from
$|\mathbf{y}| \sim 1/|\mathbf{p}_1| \ll |\mathbf{x}_1|$.
Comparing the above equations, we obtain
\begin{equation}
    |\chi_{\mathbf{k}}(\mathbf{x}_1)|^2 
    = \frac{\delta^{(3)}(\mathbf{p}_2 - \mathbf{p})}{\delta^{(3)}(\mathbf{p}_1 - \mathbf{p})}
    = J(\mathbf{p}_2, \mathbf{x}_1), 
\end{equation}
where we use the property of delta function
\begin{equation}
    \delta^{(3)}(\mathbf{p}_1 - \mathbf{p}) = \frac{1}{J(\mathbf{p}_2,\mathbf{x}_1)}\delta^{(3)}(\mathbf{p}_2 - \mathbf{p}),
\end{equation}
with $d^3 p_1 = J(\mathbf{p}_2,\mathbf{x}_1)d^3 p_2$ the Jacobian matrix that relates the infinitesimal velocity element at position $\mathbf{x}_1$ to that at asymptotic infinity.

\bibliographystyle{utphysGe}
\bibliography{ref}
\end{document}